\documentclass[conference]{IEEEtran}
\IEEEoverridecommandlockouts
\usepackage{cite}
\usepackage{amsmath,amssymb,amsfonts}
\usepackage{algorithmic}
\usepackage{graphicx}
\usepackage{textcomp}
\usepackage{xcolor}
\usepackage{color,soul,xcolor}
\usepackage{cuted}

\normalsize

\newcommand{\ket}[1]{\left|#1\right\rangle}
\newcommand{\bra}[1]{\left\langle#1\right|}

\def\BibTeX{{\rm B\kern-.05em{\sc i\kern-.025em b}\kern-.08em
    T\kern-.1667em\lower.7ex\hbox{E}\kern-.125emX}}
\begin{document}

\title{Analytic view on Coupled Single-Electron Lines\\
}

\author{\IEEEauthorblockN{1\textsuperscript{st} Krzysztof Pomorski}
\IEEEauthorblockA{\textit{University College Dublin} \\
\textit{School of Electrical and Electronic Engineering}\\
Dublin, Ireland \\
email: kdvpomorski@gmail.com}
\and
\IEEEauthorblockN{2\textsuperscript{nd} Panagiotis Giounanlis}
\IEEEauthorblockA{\textit{University College Dublin} \\
\textit{ School of Electrical and Electronic Engineering}\\
Dublin, Ireland \\
email:panagiotis.giounanlis@ucd.ie }
\and
\IEEEauthorblockN{3\textsuperscript{rd} Elena Blokhina}
\IEEEauthorblockA{\textit{University College Dublin} \\
\textit{School of Electrical and Electronic Engineering}\\
Dublin, Ireland \\
email:elena.blokhina@ucd.ie }
%\and
%\IEEEauthorblockN{4\textsuperscript{th} Andrew Mitchell}
%\IEEEauthorblockA{\textit{University College Dublin} \\
%\textit{School of Physics}\\
%Dublin, Ireland \\
%email:andrew.mitchell@ucd.ie}
\and
\IEEEauthorblockN{4\textsuperscript{th} Imran Bashir}
\IEEEauthorblockA{\textit{Equal 1} \\
California, USA \\
email:imran.bashir@equal1.com}
\and
\IEEEauthorblockN{5\textsuperscript{th} Dirk Leipold}
\IEEEauthorblockA{\textit{Equal 1} \\
California, USA \\
email:dirk.leipold@equal1.com}
\and
\IEEEauthorblockN{6\textsuperscript{th} Robert Staszewski}
\IEEEauthorblockA{\textit{University College Dublin,School of Electrical and Electronic Engineering} \\
Dublin, Ireland \\
email:robert.staszewski@ucd.ie}
}

\maketitle

\begin{abstract}
Fundamental properties of two electrostatically interacting single-electron lines (SEL) are determined from a minimalistic tight-binding model. The lines are represented by a chain of coupled quantum wells that could be implemented in a mainstream nanoscale CMOS process technology and tuned electrostatically by DC or AC voltage biases. The obtained results show an essential qualitative difference with two capacitively coupled classical electrical lines. The derived equations and their solutions prove that the two coupled SET lines can create an entanglement between electrons. The results indicate a possibility of constructing electrostatic (non-spin) coupled qubits that could be used as building blocks in a CMOS quantum computer.
\end{abstract}


%\bibliography{bib}{}

\section{Technological Motivation}

The CMOS electronic devices continue to scale down with Moore's law and now are starting to reach the fundamental limitation dictated by the fact that the electron charge is quantized \cite{Patra_2018}, \cite{Panos_2019}. Moreover, it is commonly accepted by the technologists that the use of fractional electron charge has no practical meaning. On the other hand, the use of different representation of information as by fluxons (quantized flux of magnetic field) in Rapid Single Quantum Flux electronics turns out to have its limitations from the point of view of its size, as implementation in very large scale integration circuits \cite{Compel}. In this work, we limit ourselves to the electrostatic description of an electron-electron interaction. Current cryogenic CMOS technology development opens perspectives in implementation of CMOS quantum computer \cite{Panos_2019} or use of cryogenic CMOS as interface to superconducting quantum circuits \cite{Patra_2018}.

Fundamentally, the electron quantum properties are captured by the Schr\"odinger equation that can be obtained in the case of a single electron in effective potential or in the case of many-electron system confined by some local potential. However, the Schr\"odinger equation in a continuous position space is not the most straightforward approach to capture all electron transport properties on discrete lattices present in various types of metamaterials that can be manufactured on large scale. In this work, we use a tight-binding approximation that can be derived from the Schr\"odinger equation \cite{TightBinding}. % [1].

\begin{figure}[h]
\centering
\includegraphics[width=1.0\linewidth]{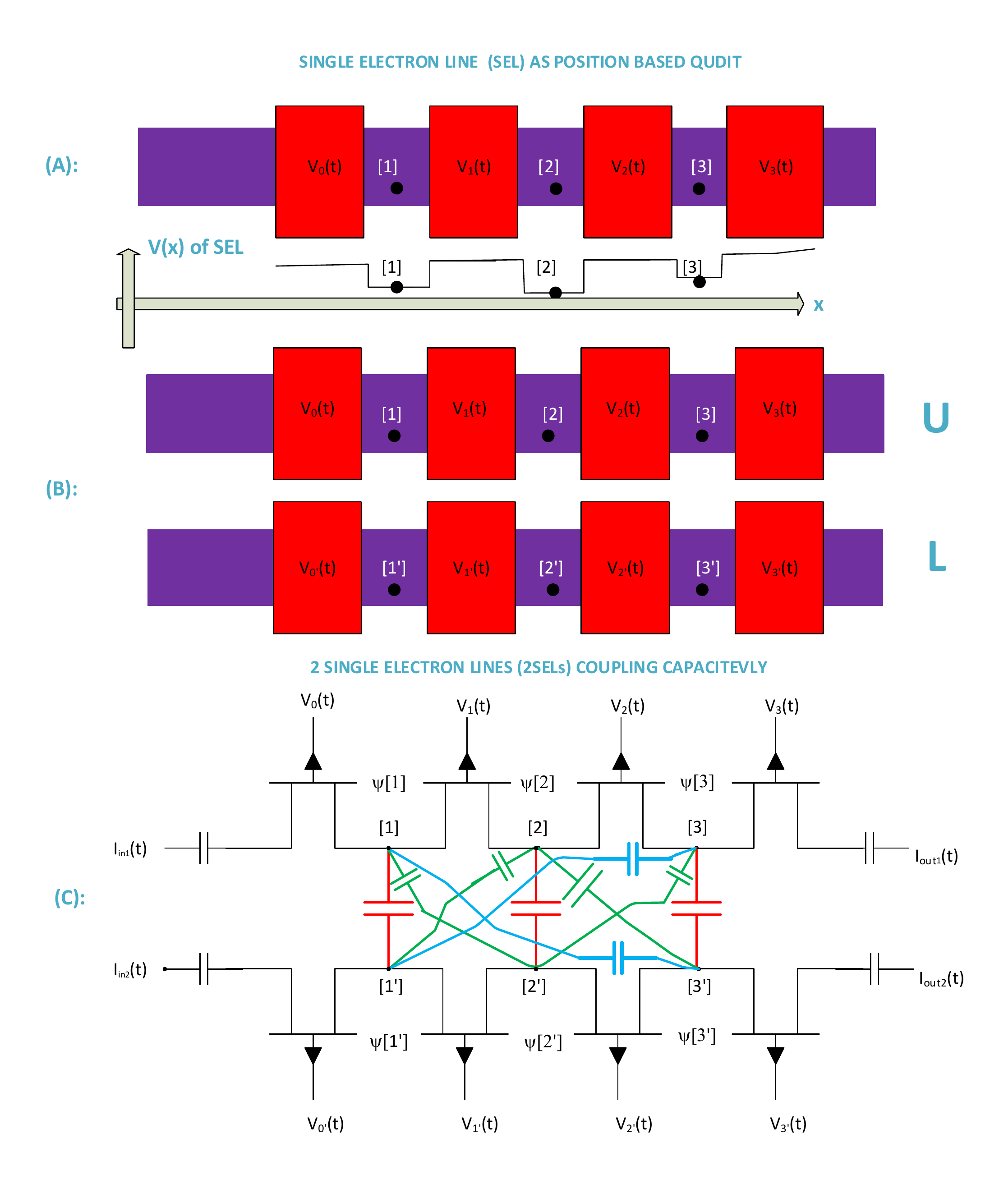} %{2SELs.png} %{2SELs_scheme.pdf} %{Transistor_SET_Lines2QQQ2.png} %{position_based_Qubit_in_RF_signal.png}
\caption{Nanometer CMOS structure, effective potential and circuit representation of: (A) electrostatic postion-dependent qudit \cite{ICECS18}; (B)--(C) two electrostatic position-dependent qudits representing two inductively interacting lines (upper "U" and lower "L" quantum systems) in minimalistic way (more rigorously they shall be named as MOS transistor single-electron lines). Presented systems are subjected to the external voltage biasing that controls the local potential landscape in which electrons are confined.  }
  \label{PositionDependentQubit}
\end{figure}

\section{Mathematical Statement of the Problem}

At first, we consider a physical system of an electron confined in a potential with two minima (position dependet qubit with presence of electron at node 1 and 2) or three minima (position dependent qudit with presence of electron at nodes 1,2 and 3), as depicted in Fig.\,\ref{PositionDependentQubit}(A), which was also considered by Fujisawa \cite{Fujisawa2004} and Petta \cite{Petta}. We can write the Hamiltonian in the second quantization as
\begin{equation}
\hat{H}=\sum_{i,j}t_{i \rightarrow j}\hat{a}^{\dag}_{i}\hat{a}_{j}+\sum_{i} E_{p}(i)\hat{a}^{\dag}_{i}\hat{a}_{i}+\sum_{i,j,k,l}\hat{a}^{\dag}_{i}\hat{a}^{\dag}_{j}\hat{a}_{i}\hat{a}_{j}V_{i,j},
\end{equation}
where $\hat{a}^{\dag}_{i}$ is a fermionic creator operator at $i$-th point in the space lattice and $\hat{a}_{j}$ is fermionic annihilator operator at $j$-th point of the lattice. The hopping term $t_{i \rightarrow j}$ describes hopping from $i$-th to $j$-th lattice point and is a measure of kinetic energy. The potential $V_{i,j}$ represents particle-particle interaction and term $E_{pi}$ incorporates potential energy. In this approach we neglect the presence of a spin.
It is convenient to write a system Hamiltonian in spectral form as
\begin{eqnarray}
\label{eq:1}
\hat{H}(t) = E_{p1}(t) \ket{1,0} \bra{1,0} +E_{p2}(t)\ket{0,1} \bra{0,1} + \nonumber \\
t_{1 \rightarrow 2}(t) \ket{0,1} \bra{1,0}+t_{2 \rightarrow 1}(t) \ket{1,0} \bra{0,1}= \nonumber \\
=\frac{1}{2}(\hat{\sigma}_0+\hat{\sigma}_3)E_{p1}(t)+\frac{1}{2}(\hat{\sigma}_0-\hat{\sigma}_3)E_{p2}(t)+ \nonumber \\
\frac{1}{2}(\hat{\sigma}_1-i\hat{\sigma}_2)t_{2 \rightarrow 1}(t)+\frac{1}{2}(i\hat{\sigma}_2-\hat{\sigma}_1)t_{1 \rightarrow 2}(t) %%%= \nonumber \\
%%=
%%\begin{pmatrix}
%%E_{p1}(t) & t_{2 \rightarrow 1} \\
%%t_{1 \rightarrow 2} & E_{p2}(t)
%%\end{pmatrix},
\end{eqnarray}
where Pauli matrices are $\hat{\sigma}_0, .. , \hat{\sigma}_3$ while system quantum state is given as $\ket{\psi(t)}=\alpha(t) \ket{1,0} + \beta(t) \ket{0,1}$ with $|\alpha|^2 + |\beta|^2 =1$ and is expressed in Wannier function eigenbases $\ket{1,0}=w_L(x)$ and $\ket{0,1}=w_R(x)$ which underlines the presence of electron on the left/right side as equivalent to picture from Schr\"odinger equation \cite{ICECS18}.). We obtain two energy eigenstates %\newline % %\nonumber \\
\begin{eqnarray*}
\ket{E_{1(2)}}=
\begin{pmatrix}
\frac{ (E_{p2}-E_{p1}) \pm \sqrt{4t_{1 \rightarrow 2}t_{2 \rightarrow1} + |E_{p1}-E_{p2}|^2 } }{2t_{1 \rightarrow 2}} \\
1
\end{pmatrix}=\nonumber \\
=\frac{ (E_{p2}-E_{p1}) \pm \sqrt{4t_{1 \rightarrow 2}t_{2 \rightarrow1} + |E_{p1}-E_{p2}|^2 } }{2t_{1 \rightarrow 2}}\ket{1,0}+\ket{0,1}.
\end{eqnarray*}
and energy eigenvalues of position dependent qubit % and orthogonal eigenstates
%We have 2 different eigenenergy values (E1,E2) given as
%
\begin{eqnarray}
E_{1(2)} = \frac{1}{2}(E_{p1} + E_{p2} \pm \sqrt{4t_{1 \rightarrow 2}t_{2 \rightarrow1} + |E_{p1}-E_{p2}|^2 }). %  \nonumber \\ .
\end{eqnarray}
%The base $\ket{1,0} (or \ket{0,1})$ underlines the presence of electron on the left (or right) side.
%We immediately recognize that $\ket{1,0} (\ket{0,1})$ corresponds to Wannier function of the left (right) well $w_L(x)$ $(w_R(x))$ \cite{ICECS18}. % corresponds to Wannier function of the right well.
The eigenstate depends on an external vector potential source acting on the qubit by means of $t_{1 \rightarrow 2}=|t_{1 \rightarrow 2}|e^{i \alpha}=t_{2 \rightarrow 1}^{*}$. Since every energy eigenstate is spanned by $\ket{0,1}$ and $\ket{1,0}$, we will obtain oscillations of occupancy between two wells \cite{ICECS18},\cite{Panos_2019}. It is worth-mentioning that the act of measurement will affect the qubit quantum state. Since we are dealing with a position-based qubit, we can make measurement of the electron position with the use an external single-electron device (SED) in close proximity to the qubit. This will require the use of projection operators that represent eigenenergy measurement as $\ket{E_{0(1)}}\bra{E_{0(1)}}$ or, for example, measurement of the electron position at left side so we use the projector $\ket{1,0}\bra{0,1}$.
We can extend the model for the case of three (and more) coupled wells. In such a case, we obtain the system Hamiltonian for a position based qudit: % of the following shape:
\begin{eqnarray}
  \label{simple_equation1}
  \hat{H} = \sum_{s}E_{ps}\ket{\textbf{s}}\bra{\textbf{s}}+\sum_{l,s,s \neq l}t_{s \rightarrow l}\ket{\textbf{l}}\bra{\textbf{s}},
  \end{eqnarray}
where $\ket{\textbf{1}}=\ket{1,0,0}, \ket{\textbf{2}}=\ket{0,1,0}, \ket{\textbf{3}}=\ket{0,0,1} $
%%\begin{eqnarray}
%%  \label{simple_equation1}
%%  \hat{H} = E_{p1} \ket{1,0,0} \bra{1,0,0} + E_{p2} \ket{0,1,0} \bra{0,1,0} + \nonumber \\
%%  E_{p3} \ket{0,0,1} \bra{0,0,1} + t_{1 \rightarrow 2} \ket{0,1,0} \bra{1,0,0} \nonumber \\
%%  +t_{2 \rightarrow 1} \ket{1,0,0} \bra{0,1,0} + t_{2 \rightarrow 3} \ket{0,0,1} \bra{0,1,0}+ \nonumber \\ t_{3 \rightarrow 2} \ket{0,1,0} \bra{0,0,1} +t_{3 \rightarrow 1} \ket{1,0,0} \bra{0,0,1} \nonumber \\
%%  +t_{1 \rightarrow 3} \ket{0,0,1} \bra{1,0,0}.
%%\end{eqnarray}
%
and its Hamiltonian matrix and quantum state $\ket{\psi}$ of position dependent qudit (with a normalization condition $|\alpha|^2+|\beta|^2+|\gamma|^2=1$) is given as
\begin{eqnarray*}
H(t)=
\begin{pmatrix}
  E_{p1}(t) & t_{2 \rightarrow 1}(t) & t_{3 \rightarrow 1}(t) \\
  t_{1 \rightarrow 2}(t) & E_{p2}(t) & t_{3 \rightarrow 2}(t) \\
  t_{1 \rightarrow 3}(t) & t_{2 \rightarrow 3}(t) & E_{p3}(t)
\end{pmatrix} ,
\\
\ket{\psi} =
\begin{pmatrix}
  \alpha(t) \\
  \beta(t) \\
  \gamma(t)
\end{pmatrix}
= \alpha(t) \ket{1,0,0} + \beta(t) \ket{0,1,0} + \gamma(t) \ket{0,0,1}.
\end{eqnarray*}
Coefficients $\alpha(t)$, $\beta(t)$ and $\gamma(t)$ describe oscillations of occupancy of one electron at wells 1, 2 and 3.
The problem of qudit equations of motion can be formulated by having
$\ket{\psi} = c_1(0)e^{-\frac{i}{\hbar}t E_1}\ket{E_1}+c_2(0)e^{-\frac{i}{\hbar}t E_2}\ket{E_2}+c_3(0)e^{-\frac{i}{\hbar}t E_3}\ket{E_3}$, where $|c_1(0)|^2$,$|c_2(0)|^2$ and $|c_3(0)|^2$ are probabilities of occupancy of $E_1$, $E_2$ and $E_3$ energetic levels. Energy levels are roots of 3rd order polynomial
\begin{eqnarray}
(-E_{p1} E_{p2} E_{p3} + E_{p3} t_{12}^2 + E_{p1} t_{23}^2 + (E_{p1} E_{p2} + E_{p1} E_{p3} \nonumber \\  + E_{p2} E_{p3} - t_{12}^2 -
t_{23}^2)E -(E_{p1}+E_{p2}+E_{p3}) E^2 + E^3=0,
\end{eqnarray}
where $\ket{E_1},\ket{E_2},\ket{E_3}$ are 3-dimensional  Hamiltonian eigenvectors.

By introducing two electrostatically interacting qudits, we are dealing with the Hamiltonian of the upper and lower lines as well as with their Coulomb electrostatic interactions. We are obtaining the Hamiltonian in spectral representation acting on the product of Hilbert spaces in the form of
\begin{eqnarray}
  \label{simple_equation2}
  \hat{H} = \hat{H}_u \times I_{l}+I_{u} \times \hat{H}_l+\hat{H}_{u-l}
\end{eqnarray}
where $H_u$ and $H_l$ are Hamiltonians of separated upper and lower qudits, $H_{l-u}$ is a two-line Coulomb interaction and $I_{u(l)}=\ket{1,0,0}_{u(l)}\bra{1,0,0}_{u(l)}+\ket{0,1,0}_{u(l)}\bra{0,1,0}_{u(l)}+\ket{0,0,1}_{u(l)}\bra{0,0,1}_{u(l)}$. The electrostatic interaction is encoded in $\textcolor{black}{E_c(1,1')=E_c(2,2')=E_c(3,3')=\frac{e^2}{4\pi \epsilon_0 \epsilon d}=q_1}$ (red capacitors of Fig.1) and \textcolor{black}{$q_2=E_c(2,1')=E_c(2,3')=E_c(1,2')=E_c(3,2')=\frac{e^2}{4\pi \epsilon_0 \epsilon \sqrt{d^2+(a+b)^2}}$}
%\begin{equation}
% \textcolor{blue}{q_2=\frac{q^2}{4\pi \epsilon_0 \epsilon \sqrt{d^2+(a+b)^2}}}
% \end{equation}
and electrostatic energy of green capacitors of Fig.1. is
\begin{equation}
\textcolor{black}{E_c(1,3')=E_c(3,1')=q_2=\frac{e^2}{4\pi \epsilon_0 \epsilon \sqrt{d^2+4(a+b)^2}}},
\end{equation}
where $a$, $b$ and $d$ are geometric parameters of the system, e is electron charge and $\epsilon$ is a relative dielectric constant of the material; $\epsilon_0$ corresponds to the dielectric constant of vacuum. The very last Hamiltonian corresponds to the following quantum state $\ket{\psi(t)}$ ($|\gamma_1(t)|^2+..|\gamma_9(t)|^2=1$) given as
\begin{eqnarray}
\label{stateSEL}
\ket{\psi(t)}=\gamma_1(t)\ket{1,0,0}_u\ket{1,0,0}_l+\gamma_2(t)\ket{1,0,0}_u\ket{0,1,0}_l \nonumber \\
+\gamma_3(t)\ket{1,0,0}_u\ket{0,0,1}_l +\gamma_4(t)\ket{0,1,0}_u\ket{1,0,0}_l \nonumber \\
+\gamma_5(t)\ket{0,1,0}_u\ket{0,1,0}_l + \gamma_6(t)\ket{0,1,0}_u\ket{0,0,1}_l \nonumber \\
+ \gamma_7(t)\ket{0,0,1}_u\ket{0,0,1}_l +\gamma_8(t)\ket{0,0,1}_u\ket{0,1,0}_l \nonumber \\
+\gamma_9(t)\ket{0,0,1}_u\ket{0,0,1}_l,
\end{eqnarray}
where $|\gamma_1(t)|^2$ is the probability of finding two electrons at nodes 1 and 1' at time $t$ (since $\gamma_1$ spans $\ket{1,0,0}_u\ket{1,0,0}_l$), etc. The Hamiltonian has nine eigenenergy solutions that are parametrized by geometric factors and hopping constants $t_{k,m}$ as well as energies $E_p(k)$ for the case of `u' or 'l' system. Formally, we can treat $E_{pk}=t_{k \rightarrow k} \equiv t_{k,k} \equiv t_k \in \textbf{R}$ as a hopping from $k$-th lattice point to the same lattice point $k$. We obtain the following Hamiltonian
\newpage
\vspace{-3mm}
\begin{strip}
\begin{eqnarray}
\hat{H}=
\begin{pmatrix}
\xi_{1,1'} & \textcolor{red}{t_{ 1' \rightarrow 2' }} & \textcolor{orange}{t_{1' \rightarrow 3'}} & \textcolor{green}{t_{1 \rightarrow 2}} & 0 & 0 & \textcolor{blue}{t_{1 \rightarrow 3}} & 0 & 0 \\ % & 0 & 0 & 0 & 0 & 0 \\
\textcolor{red}{t_{ 2' \rightarrow 1' }} & \xi_{1,2'} & \textcolor{gray}{t_{2' \rightarrow 3'}} & 0 & \textcolor{green}{t_{1 \rightarrow 2}} & 0 & 0 & \textcolor{blue}{t_{1 \rightarrow 3}} & 0 \\ %& 0 & 0 & 0 & 0 & 0 \\
\textcolor{orange}{t_{3' \rightarrow 1'}} & \textcolor{gray}{t_{3' \rightarrow 2'}} & \xi_{1,3'} & 0 & 0 & \textcolor{green}{t_{1 \rightarrow 2}} & 0 & 0 & \textcolor{blue}{t_{1 \rightarrow 3}} \\
\textcolor{green}{t_{2 \rightarrow 1}} & 0 & 0 & \xi_{2,1'} & \textcolor{red}{ t_{1' \rightarrow 2'} } & \textcolor{orange}{t_{1' \rightarrow 3'}} & \textcolor{yellow}{t_{2 \rightarrow 3}} & 0 & 0 \\
0 & \textcolor{green}{t_{2 \rightarrow 1}} & 0 & \textcolor{red}{t_{2' \rightarrow 1'}} & \xi_{2,2'} & \textcolor{gray}{t_{2' \rightarrow 3'}} & 0 & \textcolor{yellow}{t_{2 \rightarrow 3}} & 0 \\
0 & 0 & \textcolor{green}{t_{2 \rightarrow 1}} & \textcolor{orange}{t_{3' \rightarrow 1'}} & \textcolor{gray}{t_{3' \rightarrow 2'}} & \xi_{2,3'} & 0 & 0 & \textcolor{yellow}{t_{2 \rightarrow 3}} \\
\textcolor{blue}{t_{3 \rightarrow 1}} & 0 & 0 & \textcolor{yellow}{t_{3 \rightarrow 2}} & 0 & 0 & \xi_{3,1'} & \textcolor{red}{t_{1' \rightarrow 2'}} & \textcolor{orange}{t_{1' \rightarrow 3'}} \\
0 & \textcolor{blue}{t_{3 \rightarrow 1}} & 0 & 0 & \textcolor{yellow}{t_{3 \rightarrow 2}} & 0 & \textcolor{red}{t_{2' \rightarrow 1'}} & \xi_{3,2'} & \textcolor{gray}{t_{2' \rightarrow 3'}} \\
0 & 0 & \textcolor{blue}{t_{3 \rightarrow 1}} & 0 & 0 & \textcolor{yellow}{t_{3 \rightarrow 2}} & \textcolor{orange}{t_{3' \rightarrow 1'}} & \textcolor{gray}{t_{3' \rightarrow 2'}} & \xi_{3,3'} \\
\end{pmatrix}
=
\begin{pmatrix}
H(1)_{1',3'} & \textcolor{green}{H_{1,2}} & \textcolor{blue}{H_{1,3}} \\
\textcolor{green}{H(1)_{2,1}} & H(2)_{1',3'} & \textcolor{yellow}{H_{2,3}} \\
\textcolor{blue}{H_{3,1}} & \textcolor{yellow}{H_{3,2}} & H(3)_{1,3'} \\
\end{pmatrix}
\end{eqnarray}
\end{strip}
\vspace{-4mm}
%\end{align}
%\end{figure*}
\normalsize

\noindent with diagonal elements $ ([\xi_{1,1'}, \xi_{1,2'}, \xi_{1,3'} ]$ , $ [\xi_{2,1'}, \xi_{2,2'}, \xi_{2,3'}]$ , $ [\xi_{3,1'}, \xi_{3,2'}, \xi_{3,3'} ])$ set to
$([(E_{p1}+E_{p1'}+E_c(1,1'))$, $ (E_{p1}+E_{p2'}+E_c(1,2'))$ , $ (E_{p1}+E_{p3'} + E_c(1,3'))]$, $[((E_{p1}+E_{p1'}+E_c(1,1'))$, $ (E_{p2}+E_{p2'}+E_c(2,2'))$ , $ (E_{p2}+E_{p3'} + E_c(2,3'))]$, $[((E_{p3}+E_{p1'}+E_c(3,1'))$, $ (E_{p3}+E_{p2'}+E_c(3,2'))$, $ (E_{p3}+E_{p3'} + E_c(3,3'))])$. In the absence of magnetic field, we have $t_{k \rightarrow m}=t_{m \rightarrow k}=t_{k,l}=t_{m,k} \in \textbf{R}$ and in the case of nonzero magnetic field $t_{k,m}=t_{m,k}^{*}\in \textbf{C}$.
It is straightforward to determine the matrix of two lines with $N$ wells [=3 in this work] each following the mathematical structure of two interacting lines with three wells in each line. Matrices $H_{1,2},H_{2,3},H_{1,3}$ are diagonal of size $N \times N$ with all the same terms on the diagonal. At the same time, block matrices $H(1)_{1',N'}$,..,$H(N)_{1',N'} $ have only different diagonal terms corresponding to $((\xi_{1,N'}, .. , \xi_{1,N'} )$, .., $(\xi_{N,N'}, .. , \xi_{N,N'} ))$ elements. In simplified considerations we can set $t_{1 \rightarrow N}=t_{N \rightarrow 1}$ and $t_{1' \rightarrow N'}=t_{N' \rightarrow 1'}$ to zero since a probability for the wavefunction transfer from 1st to $N$-th lattice point is generally proportional to $\approx \exp(-s N),$ where $s$ is some constant. It shall be underlined that in the most general case of two capacitevly coupled symmetric SELs with three wells each (being parallel to each other), we have six (all different $E_{pk}$ and $E_{pl'}$) plus 6 (all different $t_{k \rightarrow s}$, $t_{k' \rightarrow s'}$) plus three geometric parameters ($d$, $a$ and $b$) as well as a dielectric constant hidden in the effective charge of interacting electrons $q$. Therefore, the model Hamiltonian has 12+4 real-valued parameters (4 depends on the material and geomtry of 2 SELs). They can be extracted from a particular transistor implementation of two SELs (Fig.\,\ref{PositionDependentQubit}C).

\section{Analytical and Numerical Modeling of Capacitevly Coupled SELs}

\subsection{Analytical Results}

The greatest simplification of matrix 8 is when we set all $t_{k' \rightarrow m'}=t_{o \rightarrow m}$ = $|t|$, and all $E_{p}(k)=E_{p}(m')=E_p$ for $N$=3. Let us first consider the case of two insulating lines (all wells on each line are completely decoupled so there is no electron tunneling between the barriers and the barrier energies are high) where there are trapped electrons so $|t|=0$ (electrons are confined in quantum wells and cannot move towards neighbouring wells). In such a case, we deal with a diagonal matrix that has three different eigenvalues on its diagonal and has three different eigenenergy values
%\[
\begin{eqnarray}
\label{InsulatorEnergy}
\hat{E}=
\left\{
  \begin{array}{lr}
E_1= q_1 = E_p + \frac{e^2}{4 \pi \epsilon \epsilon_0 d}, \\
E_2= q_2 =E_p + \frac{e^2}{4 \pi \epsilon \epsilon_0 \sqrt{|d|^2+(a+b)^2}}, \\
E_3= q_3 =E_p + \frac{e^2}{4 \pi \epsilon \epsilon_0 \sqrt{|d|^2+4(a+b)^2}}, 
  \end{array}
\right.
\end{eqnarray}
%\]

%\begin{equation}
%E_1= q_1 = E_p + \frac{q^2}{4 \pi \epsilon \epsilon_0 d},
%\end{equation}
%\begin{equation}
%E_2= q_2 =E_p + \frac{q^2}{4 \pi \epsilon \epsilon_0 \sqrt{|d|^2+(a+b)^2}},
%\end{equation}
%\begin{equation}
%E_3= q_3 =E_p + \frac{q^2}{4 \pi \epsilon \epsilon_0 \sqrt{|d|^2+4(a+b)^2}},
%\end{equation}
so $E_3<E_2<E_1$. In the limit of infinite distance between SELs, we have nine degenerate eigenergies. They are set to $E_{pk}$ which corresponds to six decoupled quantum systems (the first electron is delocalized into three upper wells, while the second electron is delocalized into three lowers wells).

Let us also consider the case of ideal metal where electrons are completely delocalized. In such a case, all $t_{k \rightarrow l (k' \rightarrow l' ) }$ $\gg$ $E_{p}( l (s') )$ 
which brings Hamiltonian diagonal terms to be negligible in comparison with other terms. In such a case, we can set all diagonal terms to be zero which is an equivalent to the case of infinitely spaced SELs lines. It simply means that in the case of ideal metals, two lines are not `seeing' each other.

Let us now turn to the case where processes associated with hopping between wells have similar values of energy to the energies denoted as $E_{p}(k(l'))$. In such a case, the Hamiltonian matrix can be parametrized only by three real value numbers due to symmetries depicted in Fig.\,\ref{PositionDependentQubit}B (we divide the matrix by a constant number $|t|$) so
%$
%\[
\begin{eqnarray}
\left\{
  \begin{array}{lr}
q_{1_1}=\frac{2E_p+\frac{e^2}{d}}{|t|}, \nonumber \\
q_{1_2}=\frac{2E_p+\frac{e^2}{\sqrt{d^2+(a+b)^2}}}{|t|}, \nonumber \\
q_{1_3}=\frac{2E_p+\frac{e^2}{\sqrt{d^2+4(a+b)^2}}}{|t|}. \nonumber
  \end{array}
\right.
\end{eqnarray}
%\]
%$
%\begin{eqnarray}
%q1_1=\frac{2E_p+\frac{q^2}{d}}{|t|},
%\end{eqnarray}
%\begin{eqnarray}
%q1_2=\frac{2E_p+\frac{q^2}{\sqrt{d^2+(a+b)^2}}}{|t|},
%\end{eqnarray}
%\begin{eqnarray}
%q1_3=\frac{2E_p+\frac{q^2}{\sqrt{d^2+4(a+b)^2}}}{|t|}.
%\end{eqnarray}
For a fixed $|t|$, we change the distance $d$ and observe that $q_{1_1}$ can be arbitrary large, while $q_{1_2}$ and $q_{1_3}$ have finite values for $d$=0. Going into the limit of infinite distance $d$, we observe that all $q_{1_1}$, $q_{1_2}$ and $q_{1_3}$ approach a finite value $\frac{2E_p}{|t|}$. We obtain the simplified Hamiltonian matrix that is a Hermitian conjugate and has a property $H_{k,k}=H_{N-k+1,N-k+1}$. It is in the form
\begin{equation}
\label{Matrix}
\hat{H}=
\begin{pmatrix}
q_{1_1} & 1 & 0 & 1 & 0 & 0 & 0 & 0 & 0 \\
1 & q_{1_2} & 1 & 0 & 1 & 0 & 0 & 0 & 0 \\
0 & 1 & q_{1_3} & 0 & 0 & 1 & 0 & 0 & 0 \\
1 & 0 & 0 & q_{1_2} & 1 & 0 & 1 & 0 & 0 \\
0 & 1 & 0 & 1 & q_{1_1} & 1 & 0 & 1 & 0 \\
0 & 0 & 1 & 0 & 1 & q_{1_2} & 0 & 0 & 1 \\
0 & 0 & 0 & 1 & 0 & 0 & q_{1_3} & 1 & 0 \\
0 & 0 & 0 & 0 & 1 & 0 & 1 & q_{1_2} & 1 \\
0 & 0 & 0 & 0 & 0 & 1 & 0 & 1 & q_{1_1} \\
\end{pmatrix}
\end{equation}

We can analytically find nine energy eigenvalues and they correspond to the entangled states. We have
\begin{equation}
\label{eigenenergies}
%\[
\left\{
  \begin{array}{lr}
E_1=q_{1_1}, \\
E_2=q_{1_2}, \\
E_3=\frac{1}{2}( q_{1_1}+q_{1_2} -\sqrt{8+(q_{1_1}-q_{1_2})^2} ), \\
E_4=\frac{1}{2}( q_{1_1}+q_{1_2} + \sqrt{8+(q_{1_1}-q_{1_2})^2} ), \\
E_5=\frac{1}{2}( q_{1_2}-q_{1_3} - \sqrt{8+(q_{1_2}-q_{1_3})^2} ), \\
E_6=\frac{1}{2}( q_{1_2}-q_{1_3} + \sqrt{8+(q_{1_2}-q_{1_3})^2} ).
  \end{array}
\right.
%\]
\end{equation}
%
%\begin{equation}
%E_1=q_{1_1}, E_2=q_{1_2}, E_3=\frac{1}{2}( q_{1_1}+q_{1_2} -\sqrt{8+(q_{1_1}-q_{1_2})^2} ),
%\end{equation}
%\begin{equation}
%E_4=\frac{1}{2}( q_{1_1}+q_{1_2} + \sqrt{8+(q_{1_1}-q_{1_2})^2} ),
%\end{equation}
% \begin{equation}
% E_5=\frac{1}{2}( q_{1_2}-q_{1_3} - \sqrt{8+(q_{1_2}-q_{1_3})^2} ),
% \end{equation}
%\begin{eqnarray}
%  E_6=\frac{1}{2}( q_{1_2}-q_{1_3} + \sqrt{8+(q_{1_2}-q_{1_3})^2} ).
%\end{eqnarray}
%
The last 3 energy eigenvalues are the most involving analytically and are the roots of a 3rd order polynomial
\begin{eqnarray*}
     2q_{1_1}+6q_{1_3}-q_{1_1} q_{1_2} q_{1_3} +(-8+q_{1_1} q_{1_2}+q_{1_1} q_{1_3}+q_{1_2} q_{1_3})E_k  \nonumber \\
      -(q_{1_1}+q_{1_2}+q_{1_3})E_k^2+E_k^3=0 \nonumber \\ .
\end{eqnarray*}
   %$2q_{1_1}+6q_{1_3}-q_{1_1} q_{1_2} q_{1_3} +(-8+q_{1_1} q_{1_2}+q_{1_1} q_{1_3}+q_{1_2} q_{1_3})E_k -(q_{1_1}+q_{1_2}+q_{1_3})E_k^2+E_k^3=0$.
We omit writing direct and very lengthy formulas since the solutions of a 3rd-order polynomial are commonly known. The eigenvectors have the structure given in Appendix.
 %%\begin{eqnarray}

We can readily recognize that all nine energy eigenvectors are entangled. In particular, $\ket{E_1}=\ket{1,0,0}_u\ket{1,0,0}_l-\ket{0,1,0}_u\ket{0,1,0}_l+\ket{0,0,1}_u\ket{0,0,1}_l$ or
 $\ket{E_2}=\ket{1,0,0}_u\ket{0,1,0}_l-\ket{0,1,0}_u\ket{1,0,0}_l - \ket{0,1,0}_u\ket{0,0,1}_l +\ket{0,0,1}_u\ket{0,1,0}_l$, so they have no equivalence in the classical picture of two charged balls in channels that are repelling each other.

\subsection{Numerical Results for Case of Capacitively Coupled SETs}

\begin{figure} %[t]
%%%\label{EnergyEigenspectra}
\centering
\includegraphics[width=0.8\linewidth]{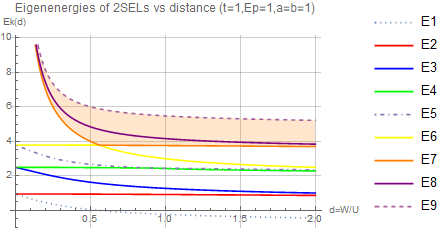} %%%InsulatorToMetal.pdf}
\includegraphics[width=0.8\linewidth]{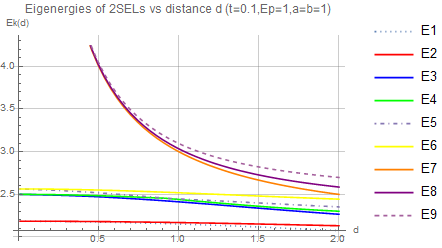}
\includegraphics[width=0.8\linewidth]{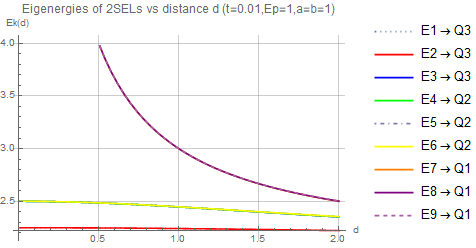}
\vspace{-4mm}
\caption{Cases of metal (top figure), semiconductor and insulator state (bottom figure) of 2SELs given by eigenenergy spectra as function of distance between two lines. Different hopping terms were used $t$=$ (1, 0.1, 0.01)$ for fixed $E_p=1$, a and b with $e=1$.
Bottom plots give the dependence of eigenenergy spectra on quantum well size, hopping term $t$ and $E_p$ parameter.} %%% The last bottom plot shows how we can tunne all eigenenergies by change of $E_p$ parameter. } %Characteristic narrowing of bands is observed when one moves from big towards small distance d between SELs (what can be related to the ration of W/U in Hubbard model) and it is one of the signs of transition from metalic to insulator regime (Mott-insulator phase transition \cite{Spalek}).  Last plot refers to Anderson localization of electrons and in such case energy eigenspectra are determined by formula \ref{InsulatorEnergy} and hopping terms $t$ can be completly neglected since electrons are localized in quantum well potential minimias.  }
\label{MottTransition1}
\includegraphics[width=0.8\linewidth]{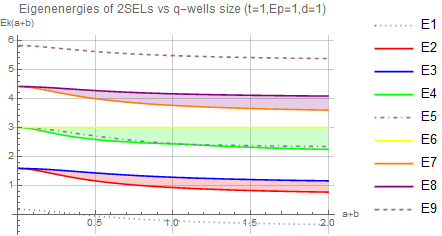}
\includegraphics[width=0.8\linewidth]{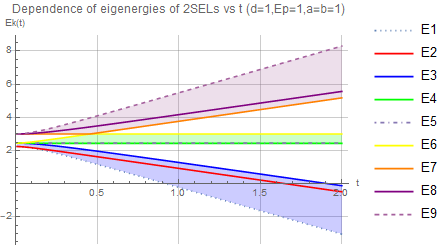}
\includegraphics[width=0.8\linewidth]{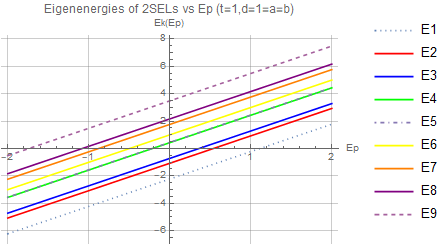}
\end{figure}

\begin{figure} %%%[b]
\centering
\includegraphics[width=0.9\linewidth]{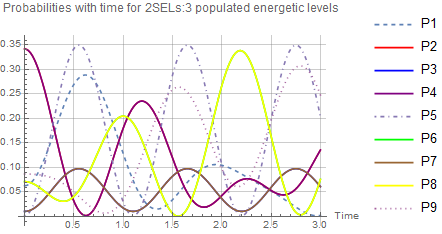}
\includegraphics[width=0.9\linewidth]{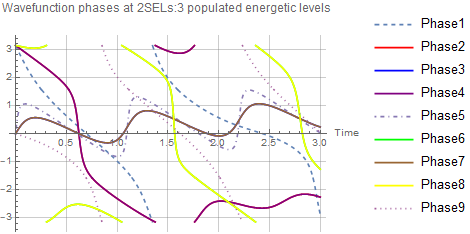}
\includegraphics[width=0.9\linewidth]{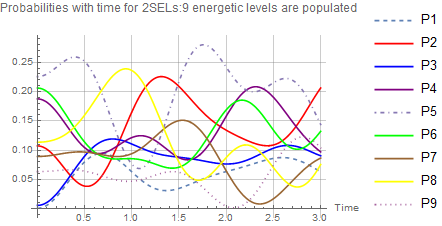}
\includegraphics[width=0.9\linewidth]{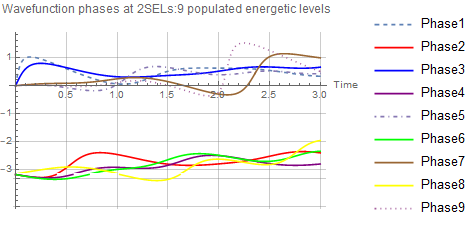}
\vspace{-4mm}
\caption{Quantum state of two SELs over time: Upper (Lower) plots populate 3 (9) energy levels. The probabilities of finding both electrons simultaneously at the input $p_1(t)=|\gamma_1(t)|^2$ and output $p_9(t)=|\gamma_1(t)|^2$ is shown with time as well as evolution of phases $\phi_1(t),..,\phi_9(t)$ of $\gamma_1(t)=|\gamma_1(t)|e^{\phi_1(t)}$, .., $\gamma_9(t)=|\gamma_9(t)|e^{\phi_9(t)}$ corresponding to equation (\ref{stateSEL}).} %% Right plots point out probabilities of finding electrons in all 9 possible configurations in 2SELs (Fig.\,\ref{PositionDependentQubit}) with time and phases of ($\gamma_1(t),..,\gamma_9(t)$).} % corresponding to 9 configurations.} % when square-root of probability occupancy for two electrons at two SELs at nodes ((1,1'),(1,2'),(1,3'),(2,1'),(2,2'),(2,3'),(3,1',(3,2'),(3,3'))) corresponds to coefficients $q_1(t)$ , .., $q_9(t)$ from the plot equivalent to $\gamma_1(t),..,\gamma_9(t)$ defined by equation \ref{stateSEL}.}
%when $$ correspond to $\gamma_1(t)$, .. , $\gamma_9(t)$ coefficients of Equation 5.}
%\caption{Transition from insulator (upper) to metallic state (bottom) given by eigenenergy spectra as function of distance between two single electron lines. Different hopping terms were used $t$= (0.01, 0.1, 1) representing continouus transition from insulator to metal  [from top to bottom plots] and $e$=1.}
%%%Two SELs over time (from top to down): Upper plots populate 3 energy levels, while other plots correspond to occupancy of 9 energy levels. Left plots describe the correlations between probabilities of finding both electrons simultaneously at the input and output. Evolution of two two-body quantum-state phases (from $\gamma_1(t)$ and $\gamma_9(t)$) is given in equation (\ref{stateSEL}). Right plots point out probabilities of finding electrons in all 9 possible configurations in 2SELs (Fig.\,\ref{PositionDependentQubit}) with time and phases of ($\gamma_1(t),..,\gamma_9(t)$).} % corresponding to 9 configurations.} % when square-root of probability occupancy for two electrons at two SELs at nodes ((1,1'),(1,2'),(1,3'),(2,1'),(2,2'),(2,3'),(3,1',(3,2'),(3,3'))) corresponds to coefficients $q_1(t)$ , .., $q_9(t)$ from the plot equivalent to $\gamma_1(t),..,\gamma_9(t)$ defined by equation \ref{stateSEL}.}
%when $$ correspond to $\gamma_1(t)$, .. , $\gamma_9(t)$ coefficients of Equation 5. }
%\caption{Transition from insulator to metallic state}
%\caption{Spectrum of eiengenergies for the case of simplified Hamiltonian in function of mobility $|t|$. }
\label{OccupancyOscillations}
\end{figure}

At first we are analyzing available spectrum of eigenenergies as in the case of insulator-to-metal phase transition \cite{Spalek}, which can be implemented in a tight-binding model by a systematic increase of the hopping term from small to large values, while at the same time keeping all other parameters constant, as depicted in Fig.\ref{MottTransition1}. Described tight-binding model can minimic  metal (t=1), semiconductor (t=0.1) and insulator state (t=0.01). We can recognized 2 SELs eigenergy spectra dependence on distance between two lines. Characteristic narrowing of bands is observed when one moves from big towards small distance d between SELs (what can be related to the ration of W/U in Hubbard model) and it is one of the signs of transition from metalic to insulator regime (Mott-insulator phase transition \cite{Spalek}). One of the plot referring to t=0.01 describes Anderson localization of electrons and in such case energy eigenspectra are determined by formula \ref{InsulatorEnergy} and hopping terms $t$ can be completly neglected since electrons are localized in quantum well potential minimias. %%%Bottom plots of Fig.2. describe the ability of tunning eigenenergy spectra by change of $E_p$ and $t$ parameter that is directly voltage controlled as one is referring to Fig.1.  
%%%It is informative to notice that change of quantum wells lenght does not affect the eigenenergy of 2SELs significantly. 
%We obtain the spectrum energies as given in Fig.2 for various cases of Ep and mobility $|t|$. transition was detected and its signature is the interchange of two first energetic levels. %We have reported various energy switching subjected to tunnable parameter that is the electron mobility $|t|$ or $E_p$ or distance. However for certain set of parameters the quantum system shows ordinary behaviour and no phase transitions are observed.
 %%%InsulatorToMetal.pdf}
%\newpage
Bottom plots of Fig. \ref{MottTransition1}. describe the ability of tunning eigenenergy spectra by change quantum well lenghts (a+b), $E_p$ and $t$ parameter. Last two parameters are directly voltage controlled as one is referring to Fig.\ref{PositionDependentQubit}, where 8 voltage signals are used for controlling effective tight-binding Hamiltonian. It is informative to notice that change of quantum wells lenght expressed by a+b does not affect the eigenenergy of 2SELs significantly. Observed change changes the ratio of electrostatic to kinetic energy and thus is similar to the change in energy eigenspectra generated by different distances d. We can spot narrowing bands when we are moving from the situation of lower to the situation higher electrostatic energy of interacting electron and again it it typical for metal-insulator phase transition. Change of ratio kinetic to electrostatic energy can be obtained by keeping quantum well size constant, constant distance between 2 SELs and by change of hopping constant $t$ that is the measure of electron ability in conducting electric or heat current. Again one observes the narrowing of bands when we reduce t so the dominant energy of electron is due to electron-electron interaction. The last plot of Fig.\ref{MottTransition1} describes our ability of tunning eigenenergy spectra of system in linear way just by change of $E_p$ parameter. In very real way we can recognize the ability of tunning the chemical potential (equivalent to Fermi energy at temperatures T=0K) by controlling voltages given in Fig.\ref{PositionDependentQubit}. in our artifcial lattice system. Due to controllability of energy eigenspectra by controlling voltages from Fig.\ref{PositionDependentQubit} one can recognize 2 SELs system as the first stage of implementation of programmable quantum matter. In general case considered 2SELs Hamiltonian consists 6 different $E_p$ parameters and 6 different $t$ parameters that can be controlled electrostatically (12 parameters under electrostatic control) by 2SELS controlling voltages $V_0(t), .., V_3(t), V_{0'}(t), .., V_{3'}(t) $ depicted in Fig.1.  \newline \newline The numerical modeling of electron transport across coupled SELs is about solving a set of nine coupled recurrent equations of motion as it is in the case of time-dependent 2 SELs Hamiltonian.
In this work we consider time-independent Hamiltonian implying constant occupation of energetic levels. Therefore the quantum state can be written in the form
$\ket{\psi(t')}$ $=\alpha_1 e^{\frac{\hbar}{i}E_1 t'}\ket{E_1}+$..$+\alpha_9 e^{\frac{\hbar}{i}E_9 t'}\ket{E_9}$, so the probability of occupancy of energetic level $E_1$ is $|\alpha_1|^2=|\bra{E_1} \ket{\psi(t)}|^2=p_{E1}=constant$, etc. Since we have obtained analytical form of all states $\ket{E_k}$ and eigenenergies $E_k$ we have analytical form of quantum state dynamics $\ket{\psi(t')}$ with time. From obtained analytical solutions presented in the Appendix we recognize that every eigenenergy state is the linear combination of positon-based states $\ket{k} \bigotimes \ket{l'}$ what will imply that quantum state can never be fully localized at two nodes k and l' as it is pointed by analytically obtained eigenstates of the 2SELs Hamiltonian that are given in the Appendix. 
In the conducted numerical simulations we visualize analytical solutions. We set $\hbar=1$ and $\alpha_1=..=\alpha_8=\frac{1}{9}$, $\alpha_9=\sqrt{1-\frac{8}{81}}$ or $\alpha_1=\alpha_2=\frac{1}{2}$,$\alpha_9=\frac{\sqrt{2}}{2}$,$\alpha_3=..=\alpha_8$ that will correspond to top or bottom plots of Fig.\ref{OccupancyOscillations}. We can recognize that probability of occupancy of (1,1') from Fig.1. (when two electrons are at input of 2SELs)  is given by $|(\bra{1,0,0}\bigotimes\bra{1,0,0})\ket{\psi(t)}|^2=|\gamma_1(t)|^2=p_1(t)$ (two electrons as SELs inputs) can be compared with occupancy of (3,3') given by $p_9(t)=|\gamma_9(t)|^2=|(\bra{0,0,1}\bigotimes\bra{0,0,1})\ket{\psi(t)}|^2$ (2 electrons at SELs outputs) as depicted in Fig.\,3. We identify probability of finding first electron at input as the sum of $p_1(t)+p_2(t)+p_3(t)$.
Various symmetries can be traced in the bottom of Fig.\ref{OccupancyOscillations}. as between probability $p_2(t)$ and $p_8(t)$ or in the upper part of Fig.
\ref{OccupancyOscillations} when $p_2(t)=p_8(t)$ or $\phi_2(t)=phase(\gamma_2(t))=\phi_8(t)$.   The same symmetry relations applies to the case of probability $p_4(t)$ and $p_6(t)$ as well as $\phi_4(\gamma_4(t))$ and $\phi_6(\gamma_6(t))$. These symmetries has its origin in the fact that 2 SELs system is symmetric along x axes what can be recoginezed in symmetries of simplified Hamiltonian matrix \ref{Matrix}. It shall be underlined that in the most general case when system matrix has no symmetries the energy eigenspectra might have less monotonic behaviour.    
 %We can also compare the situation of measurement of only one electron at the output of each line that is given by the output of first upper SEL that is correlated with $p_7=|\gamma_7(t)|^2$ (that spans the state $\ket{0,0,1}_u\ket{1,0,0}_l$) or to $p_8=|\gamma_8(t)|^2$ (that spans the state $\ket{0,0,1}_u\ket{0,1,0}_l$). %Alternatively, we can consider the one electron output of the second line while there is no electron at the output of the first line, which corresponds to $\gamma_3(t)$  spanning the state $\ket{1,0,0}_u\ket{0,0,1}_l$ or to the coefficient $\gamma_6(t)$ spanning the state $\ket{0,1,0}_u\ket{0,0,1}_l$. %(In previous considerations instead of $q$ we have used $\gamma$). It is worth-mentioning that act of measurement done in eigenenergy or in position bases will change the system dynamics with time. It is not suprising since measurement is the interaction of external quantum /classical system on our investigated system.
%\onecolumn

%\vspace{-1mm}

\subsection{Act of Measurement and Dynamics of Quantum State}

The quantum system dynamics over time is expressed by the equation of motion $\hat{H}(t')\ket{\psi(t')}=i\hbar \frac{d}{dt'}\ket{\psi(t')}$ that can be represented in discrete time step by relation
%\begin{equation}
    $\frac{dt'}{i \hbar}\hat{H}(t')\ket{\psi(t')}+\ket{\psi(t')}=\ket{\psi(t'+dt')}$.
%\end{equation}
%
It leads to the following equations of motion for quantum state expressed by equation (\ref{stateSEL}) as follows
\begin{eqnarray}
%\[
\vec{\gamma}(t'+dt')=
\left\{
  \begin{array}{lr}
\gamma_1(t')+dt'\sum_{k=1}^{9} \hat{H}_{1,k}(t')\gamma_k(t')=\\ f_1(\vec{\gamma}(t'),dt')[\hat{H}(t')], \\
%%\gamma_2(t'+dt')=\gamma_2(t')+dt', \\
.. \\
\gamma_9(t')+dt'\sum_{k=1}^{9} \hat{H}_{9,k}(t')\gamma_k(t')=\\ f_9(\vec{\gamma}(t'),dt')[\hat{H}(t')] \\
%%\gamma_9(t'+dt')=\gamma_9(t')+dt'.
  \end{array}
\right\}= \nonumber\\ =\vec{f}(\vec{\gamma}(t'),dt')[\hat{H}(t')]=\vec{f}(\vec{\gamma}(t'),dt')_{[\hat{H}(t')]}. \nonumber \\
%\]
\end{eqnarray}
%
%\onecolumn
Symbol $[.]$ denotes functional dependence of $\vec{f}(\vec{\gamma}(t'),dt')$ on Hamiltonian $\hat{H}(t')$. The measurement can be represented by projection operators $\hat{\Pi}(t')$ equivalent to the matrix that acts on the quantum state over time. The lack of measurement can simply mean that the state projects on itself so the projection is the identity operation ($\hat{\Pi}(t')=\hat{I}_{9 \times 9}$). Otherwise, the quantum state is projected on its subset and hence the projection operator can change in a non-continuous way over time. We can formally write the quantum state dynamics with respect to time during the occurrence of measurement process (interaction of external physical system with the considered quantum system) as
%
%%\begin{strip}
\begin{eqnarray*}
\vec{\gamma}(t'+dt')= \nonumber \\
\frac{\hat{\Pi}(t'+dt')(\vec{f}(\vec{\gamma}(t'),dt')_{[\hat{H}(t')]})}{(\hat{\Pi}(t'+dt')\vec{f}(\vec{\gamma}(t'),dt')_{[\hat{H}(t')]})^{\dag}(\hat{\Pi}(t'+dt')\vec{f}(\vec{\gamma}(t'),dt')_{[\hat{H}(t')]})}. %\nonumber \\
\end{eqnarray*}
%%\end{strip}
%
Let us refer to some example by assuming that a particle in the upper SELs was detected by the upper output detector (Fig.\,\ref{PositionDependentQubit}b). In such a case, the following projector $\hat{\Pi}(t,t+\Delta t) $ is different from the identity in time interval $(t,t+\Delta t)$ with $1_{1_{t,t+\Delta t}}=1$ set to 1 in this time interval and 0 otherwise. The projector acts on the quantum state (diagonal matrix is given by diag symbol). It is given as
%\begin{strip}
\begin{eqnarray}
    \hat{\Pi}(t,t+\Delta t)=  (1-1_{t,t+\Delta t})(\hat{I}_U \times \hat{I}_L)+ \nonumber \\ 1_{t,t+\Delta t}(\ket{0,0,1}_U\bra{0,0,1}_U \times \hat{I}_L)= \nonumber \\
   (1-1_{t,t+\Delta t})(\hat{I}_U \times \hat{I}_L)+ \nonumber \\
1_{t,t+\Delta t}( \ket{0,0,1}_U\bra{0,0,1}_U \times (\ket{1,0,0}_L\bra{1,0,0}_L+ \nonumber \\  \ket{0,1,0}_L\bra{0,1,0}_L+ \ket{0,0,1}_L\bra{0,0,1}_L)) = \nonumber \\
   =(1-1_{t,t+\Delta t})\hat{I}_{9 \times 9}
  % \begin{pmatrix}
  %  1 & 0 & 0 & 0 & 0 & 0 & 0 & 0 & 0 \\
  %  0 & 1 & 0 & 0 & 0 & 0 & 0 & 0 & 0 \\
  %  0 & 0 & 1 & 0 & 0 & 0 & 0 & 0 & 0 \\
  %  0 & 0 & 0 & 1 & 0 & 0 & 0 & 0 & 0 \\
  %  0 & 0 & 0 & 0 & 1 & 0 & 0 & 0 & 0 \\
  %  0 & 0 & 0 & 0 & 0 & 1 & 0 & 0 & 0 \\
  %  0 & 0 & 0 & 0 & 0 & 0 & 1 & 0 & 0 \\
  %  0 & 0 & 0 & 0 & 0 & 0 & 0 & 1 & 0 \\
  %  0 & 0 & 0 & 0 & 0 & 0 & 0 & 0 & 1 \\
  % \end{pmatrix}
  %\nonumber \\
 +1_{t,t+\Delta t} diag(0,0,1) \times \hat{I}_{3 \times 3}
   %\begin{pmatrix}
   % 0 & 0 & 0 & 0 & 0 & 0 & 0 & 0 & 0 \\
   % 0 & 0 & 0 & 0 & 0 & 0 & 0 & 0 & 0 \\
   % 0 & 0 & 0 & 0 & 0 & 0 & 0 & 0 & 0 \\
   % 0 & 0 & 0 & 0 & 0 & 0 & 0 & 0 & 0 \\
   % 0 & 0 & 0 & 0 & 0 & 0 & 0 & 0 & 0 \\
   % 0 & 0 & 0 & 0 & 0 & 0 & 0 & 0 & 0 \\
   % 0 & 0 & 0 & 0 & 0 & 0 & 1 & 0 & 0 \\
   % 0 & 0 & 0 & 0 & 0 & 0 & 0 & 1 & 0 \\
   %    0 & 0 & 0 & 0 & 0 & 0 & 0 & 0 & 1 \\
   % \end{pmatrix}
 \\ \nonumber  =diag((1-1_{t,t+\Delta t}),(1-1_{t,t+\Delta t}),(1-1_{t,t+\Delta t}), \\ \nonumber
(1-1_{t,t+\Delta t}),(1-1_{t,t+\Delta t}),(1-1_{t,t+\Delta t}),1,1,1)
   %\begin{pmatrix}
%     (1-1_{t,t+\Delta t}) & 0 & 0 & 0 & 0 & 0 & 0 & 0 & 0 \\
%    0 &  (1-1_{t,t+\Delta t}) & 0 & 0 & 0 & 0 & 0 & 0 & 0 \\
%    0 & 0 &  (1-1_{t,t+\Delta t}) & 0 & 0 & 0 & 0 & 0 & 0 \\
%    0 & 0 & 0 &  (1-1_{t,t+\Delta t}) & 0 & 0 & 0 & 0 & 0 \\
%    0 & 0 & 0 & 0 &  (1-1_{t,t+\Delta t}) & 0 & 0 & 0 & 0 \\
%    0 & 0 & 0 & 0 & 0 &  (1-1_{t,t+\Delta t}) & 0 & 0 & 0 \\
%    0 & 0 & 0 & 0 & 0 & 0 & 1 & 0 & 0 \\
%    0 & 0 & 0 & 0 & 0 & 0 & 0 & 1 & 0 \\
%    0 & 0 & 0 & 0 & 0 & 0 & 0 & 0 & 1 \\
%   \end{pmatrix}.
\end{eqnarray}
\subsection{Analogies of coupled SELs with other physical systems}
The repulsion  (anticorrelation in position) of two electrons occurs in two parallel SELs and can be used in the construction of quantum SWAP gate. Therefore, the results obtained analytically and numerically on the two interacting SELs has its meaning in the development of quantum technologies \cite{ICECS18} and also point to the interlink between the fundamental and applied science.
 It is important to underline that the tight-binding model allows for a quick detection of entangled states and for a transfer of this information into Schr\"odinger formalism, which has its importance in the design of quantum computer consisting of many coupled entangled qubits (fundamental modelling due to the large Hilbert space is limited to 10 qubits). Quite obviously, the Schr\"odinger equation gives detailed space resolution of quantum mechanical phenomena taking place in 2 or $N$ electrostatically coupled SELs that might contain an arbitrary number of quantum wells. Incorporation of spin effects is also possible in the given framework, but is beyond the scope of this work.
The tight-binding model can be derived from the Schr\"odinger formalism (and vice versa) and is the simplistic version of the Hubbard model that is a very universal model capable of descring various collective phenomena in condensed matter physics.
%and is capable of describing various phase transitions, various transport properties in fermionic and bosonic systems.
Therefore, it is expected that the tight-binding model can be also effective in describing physical effects in % the Josephson junctions and
various programmable (electrostatically controlled) CMOS nanostructures. It shall be also underlined that the described position-dependent qubits are analogical to superconducting Cooper pair boxes where quantum phase transitions have been observed \cite{Choi}, \cite{QPTBelzig}.
The hopping term in the tight-binding model of semiconductor position-dependent qubits is analogical to energy of Josephson coupling in superconducting Cooper pair box (or in other types of superconducting qubits). Therefore, the existence of quantum phase transitions \cite{QPT} is expected to occur in the studied SELs system since quantum phase transitions occurs in arrays of electrostatically coupled Josephson junctions. Therefore one is expecting to spot quantum phase transition in SELs coupled to superconducting Cooper pair boxes.
%%%\twocolumn
%%\newpage
\section{Conclusion}

%as well as hybrid systems (semiconductor-superconducting) of various topologies.
Described two Single Electron lines are approximated by occupation of two electrons at three different nodes in space at each line: 1(1'), 2(2') and 3(3') as it is given in Fig.\ref{PositionDependentQubit}. In such way two electrostatic semiconductor position based qudits can be characterized what makes the result of this work valid for the case of two capacitevly interacting semiconductor qudits or qubits \cite{Panos_2019}. In the presented work, new qualitative features of two capacitevly coupled single-electron lines were described as occupancy oscillations at SELs nodes depicted in Fig.\,\ref{OccupancyOscillations}. Obtained occupancy oscillations do not occur in the classically coupled electrical lines and thus are the feature of quantumness of the studied structure. From obtained solutions we can spot the possible transitions between energy levels (depicted in Fig.\,\ref{MottTransition1} as function of distance between SELs) when the system is subjected to the external microwave field as coming from RF sources placed in the proximity of SELs what can be factor controlling physical state of coupled SELs. What is more entangled eigenstates were obtained as analytical solutions of simplified system matrix Hamiltonian $\hat{H}$  (equation \ref{Matrix}) and are given by formulas \ref{ent1}-\ref{ent7} in the Appendix. The entangled states corresponds to eigenenergies obtained analytically and given by formulas \ref{eigenenergies} and depend on SELs distance as expressed by Fig.\,\ref{MottTransition1}.
The conducted study has its relevance in single-electron transistor structures as deriving from nanoscale CMOS. Such systems are expected to mimic various types of programmable quantum matter that can simulate many types of physical phenomena as $E_p$ and $t$ parameters of tight-binding model can be controlled electrostatically in single electron transistors. One of the interesting illustration of this is the imitation of metal-insulator phase transition in coupled nanowires as given in Fig.\,\ref{MottTransition1} that can be obtained with electrical tuning of 2-SELs system. %It shall be also underlined that the described position-dependent qubits are analogical to superconducting Cooper pair boxes where quantum phase transitions have been observed \cite{Choi}, \cite{QPTBelzig}.
\section{Acknowledgment}
%\textit{
We would like to thank to Andrew Mitchell from University College Dublin, Adam Bednorz from University of Warsaw and to Amir Bozorg from University College Dublin for fruitful discussions and remarks. This work was supported by Science Foundation Ireland under Grant 14/RP/I2921.
%}
\section{Appendix}
The simplified Hamiltonian (given by equation \ref{Matrix}) for 2 electrostatically interacting Single Electron Lines (Fig.\ref{PositionDependentQubit}.) has eigevalues pointed by formulas (10)-(12) and has following eigenvectors
%
%%\begin{eqnarray}
%%\label{Matrix1}
%\begin{center}
%%\hat{H}=
%%\begin{pmatrix}
%%q_{1_1} & 1 & 0 & 1 & 0 & 0 & 0 & 0 & 0 \\
%%1 & q_{1_2} & 1 & 0 & 1 & 0 & 0 & 0 & 0 \\
%%0 & 1 & q_{1_3} & 0 & 0 & 1 & 0 & 0 & 0 \\
%%1 & 0 & 0 & q_{1_2} & 1 & 0 & 1 & 0 & 0 \\
%%0 & 1 & 0 & 1 & q_{1_1} & 1 & 0 & 1 & 0 \\
%%0 & 0 & 1 & 0 & 1 & q_{1_2} & 0 & 0 & 1 \\
%%0 & 0 & 0 & 1 & 0 & 0 & q_{1_3} & 1 & 0 \\
%%0 & 0 & 0 & 0 & 1 & 0 & 1 & q_{1_2} & 1 \\
%%0 & 0 & 0 & 0 & 0 & 1 & 0 & 1 & q_{1_1} \\
%%\end{pmatrix},
%\end{equation}
%%%\end{eqnarray}
\begin{eqnarray}
\label{ent1}
% \begin{center}
 \ket{E_1}=
 \begin{pmatrix}
 1, \\
 0, \\
 0, \\
 0, \\
 -1, \\
 0, \\
 0, \\
 0, \\
 1 \\
 \end{pmatrix},
 \ket{E_2}=
 \begin{pmatrix}
 0, \\
 1, \\
 0, \\
 -1, \\
 0, \\
 -1, \\
 0, \\
 1, \\
 0 \\
 \end{pmatrix}, %\\
%%%\end{eqnarray}
%%\begin{eqnarray}
%%\label{ent3}
 %\end{center}
%%  \ket{E_3}=
%% \begin{pmatrix}
%% -1, \\
%% \frac{1}{4}( q_{1_1} - q_{1_2} + \sqrt{8+(q_{1_1}-q_{1_2})^2} ), \\
 %%0, \\
 %%\frac{1}{4}( q_{1_1} - q_{1_2} + \sqrt{8+(q_{1_1}-q_{1_2})^2} ), \\
%% 0, \\
%% -\frac{1}{4}( q_{1_1} - q_{1_2} + \sqrt{8+(q_{1_1}-q_{1_2})^2} ), \\
 %%0, \\
 %%-\frac{1}{4}( q_{1_1} - q_{1_2} + \sqrt{8+(q_{1_1}-q_{1_2})^2} ), \\
 %%1 \\
%% \end{pmatrix},
 \end{eqnarray}
\begin{eqnarray}
\label{ent4}
  \ket{E_{3(4)}}=
 \begin{pmatrix}
 -1, \\
 \frac{1}{4}( q_{1_1} - q_{1_2} \pm \sqrt{8+(q_{1_1}-q_{1_2})^2} ), \\
 0, \\
 \frac{1}{4}( q_{1_1} - q_{1_2} \pm \sqrt{8+(q_{1_1}-q_{1_2})^2} ), \\
 0, \\
 -\frac{1}{4}( q_{1_1} - q_{1_2} \pm \sqrt{8+(q_{1_1}-q_{1_2})^2} ), \\
 0, \\
 -\frac{1}{4}( q_{1_1} - q_{1_2} \pm \sqrt{8+(q_{1_1}-q_{1_2})^2} ), \\
 1 \\
 \end{pmatrix}, \\
 %\end{eqnarray}
 %\begin{eqnarray}
\label{ent5}
   \ket{E_{5(6)}}=
 \begin{pmatrix}
 -1, \\
 \frac{1}{4}( q_{1_2} - q_{1_3} \pm \sqrt{8+(q_{1_2}-q_{1_3})^2} ), \\
 0, \\
 \frac{1}{4}( q_{1_2} - q_{1_3} \pm \sqrt{8+(q_{1_2}-q_{1_3})^2} ), \\
 0, \\
 -\frac{1}{4}( q_{1_2} - q_{1_3} \pm \sqrt{8+(q_{1_2}-q_{1_3})^2} ), \\
 0, \\
 -\frac{1}{4}( q_{1_2} - q_{1_3} \pm \sqrt{8+(q_{1_2}-q_{1_3})^2} ), \\
 1 \\
 \end{pmatrix}, %\nonumber \\
 %\end{eqnarray}
%% \begin{eqnarray}
%%\label{ent6}
%% \ket{E_6}=
%% \begin{pmatrix}
 %%-1, \\
 %%\frac{1}{4}( q_{1_2} - q_{1_3} - \sqrt{8+(q_{1_2}-q_{1_3})^2} ), \\
 %%0, \\
 %%\frac{1}{4}( q_{1_2} - q_{1_3} - \sqrt{8+(q_{1_2}-q_{1_3})^2} ), \\
 %%0, \\
%% -\frac{1}{4}( q_{1_2} - q_{1_3} - \sqrt{8+(q_{1_2}-q_{1_3})^2} ), \\
%% 0, \\
%% -\frac{1}{4}( q_{1_2} - q_{1_3} - \sqrt{8+(q_{1_2}-q_{1_3})^2} ), \\
%% 1 \\
%% \end{pmatrix},  
\end{eqnarray}
 \begin{eqnarray}
\label{ent7}
 \ket{E_{k=(7 .. 9)}}=
 \begin{pmatrix}
 1, \\
(E_{k=(7 .. 9)} - q_{1_1})/2, \\
\frac{(-E_{k=(7 .. 9)} + q_{1_1}) ( -2 + E_{k=(7 .. 9)}^2 + q_{1_1} q_{1_2} - E_{k=(7 .. 9)} (q_{1_1} + q_{1_2}))}{ 2 (-3 E_{k=(7 .. 9)} + q_{1_1} + 2 q_{1_3})}, \\
(E_{k=(7 .. 9)} - q_{1_1})/2, \\
2, \\
(E_{k=(7 .. 9)} - q_{1_1})/2 , \\
2, \\
\frac{(-E_{k=(7 .. 9)} + q1_1)(-2 + E_{k=(7 .. 9)}^2 + q_{1_1} q_{1_2} -E_{k=(7 .. 9)} (q_{1_1} + q_{1_2}))} {2 (-3 E_{k=(7 .. 9)} + q_{1_1} + 2 q_{1_3})}
\end{pmatrix}. \\
 \end{eqnarray}
%Obtained eigenstates has no classical counterpart.  
%Obtained eigenstates has no classical counterpart. 
\bibliographystyle{plain}

\end{document}